\documentclass[aoas,nameyear,seceqn,dvips]{arximspdf}
\usepackage{dcolumn}
\usepackage{graphicx}

\doi{10.1214/10-AOAS436}
\volume{5}
\issue{2B}
\pubyear{2011}
\firstpage{1512}
\lastpage{1533}

\makeatletter
\newcolumntype{d}[1]{D{.}{.}{#1}}
\makeatother

\begin{document}
\begin{frontmatter}

\title{Population size estimation based upon ratios of recapture probabilities}
\runtitle{Population size estimation based upon ratios}

\begin{aug}
\author[a]{\fnms{Irene} \snm{Rocchetti}\thanksref{t1}\ead[label=e1]{irene.rocchetti@uniroma1.it}},
\author[b]{\fnms{John} \snm{Bunge}\thanksref{t2}\ead[label=e2]{j.bunge@cornell.edu}}
\and
\author[c]{\fnms{Dankmar} \snm{B\"{o}hning}\corref{}\ead[label=e3]{d.a.w.bohning@reading.ac.uk}
\ead[label=u1,url]{http://www.reading.ac.uk/\textasciitilde sns05dab}}

\thankstext{t1}{Supported by
Sapienza University of Rome, grant ``Mixed Effect Models for
Heterogeneous Data Modeling.''}
\thankstext{t2}{This research was conducted using the resources of the
Cornell University Center for Advanced Computing, which receives
funding from Cornell University, New York State, the National Science
Foundation and other leading public agencies, foundations and
corporations. Supported by NSF Grant DEB-08-16638.}
\runauthor{I. Rocchetti, J. Bunge and D. B\"{o}hning}

\affiliation{Sapienza University of Rome, Cornell
University and University of Reading}

\address[a]{I. Rocchetti\\
Department of Demography\\
Sapienza University of Rome\\
Rome\\
Italy\\
\printead{e1}}
\address[b]{
J. Bunge\\
Department of Statistical Sciences\\
Cornell University\\
Ithaca, New York\\
USA\\
\printead{e2}}

\address[c]{D. B\"{o}hning\\
Department of Mathematics\\
\quad and Statistics\\
School of Mathematical\\
\quad and Physical Sciences\\
University of Reading\\
Reading\\
United Kingdom\\
\printead{e3}\\
\printead{u1}}
\end{aug}

\received{\smonth{11} \syear{2009}}
\revised{\smonth{10} \syear{2010}}

%
\begin{abstract}
Estimating the size of an elusive target population is
of prominent interest in many areas in the life and social sciences.
Our aim is to provide an efficient and workable method to estimate the unknown
population size, given the frequency distribution of counts of
repeated identifications of units of the population of interest.
This counting variable is necessarily zero-truncated, since units
that have never been identified are not in the sample. We consider
several applications: clinical medicine, where interest
is in estimating patients with adenomatous polyps which have been
overlooked by the diagnostic procedure; drug user studies, where
interest is in estimating the number of hidden drug users which are
not identified; veterinary surveillance of scrapie in the
UK, where interest is in estimating the hidden amount of scrapie; and
entomology and microbial ecology, where interest is in estimating the
number of unobserved species of organisms. In
all these examples, simple models such as the homogenous Poisson
are not appropriate since they do not account for present and
latent heterogeneity. The Poisson--Gamma (negative binomial) model
provides a flexible
alternative and often leads to well-fitting models. It has a long
history and was recently used in the development of the Chao--Bunge
estimator. Here
we use a different property of the Poisson--Gamma model: if we
consider ratios of neighboring Poisson--Gamma probabilities, then
these are linearly related to the counts of repeated
identifications. Also, ratios have the useful property that they are
identical for truncated and untruncated distributions. 
In this
paper we propose a weighted logarithmic regression model to estimate
the zero frequency counts, assuming a Gamma--Poisson distribution
for the counts. A detailed explanation about the chosen weights and
a goodness of fit index are presented, along with extensions to other
distributions. 
To evaluate the proposed
estimator, we applied it to the benchmark examples mentioned above,
and we compared the results with those obtained through the
Chao--Bunge and other estimators. The major benefits of the proposed
estimator are
that it is defined under mild conditions, whereas the Chao--Bunge
estimator fails to be well defined in several of the examples
presented; in cases where the Chao--Bunge estimator is defined, its
behavior is comparable to the proposed estimator in terms of Bias and MSE
as a simulation study shows. Furthermore, the proposed estimator is
relatively insensitive to inclusion or exclusion of large outlying
frequencies, while sensitivity to outliers is characteristic of most
other methods. The implications and limitations of
such methods are discussed.

\end{abstract}

%
\begin{keyword}
\kwd{Chao--Bunge estimator}
\kwd{Katz distribution}
\kwd{species problem}
\kwd{negative binomial distribution}
\kwd{weighted linear regression}
\kwd{zero-truncation}.
\end{keyword}

\end{frontmatter}
%

\section{Introduction} \label{intro}

The size $N$ of an elusive population must often be determined. Elusive
populations occur, for example, in public health and medicine,
agriculture and veterinary science, software engineering, illegal
behavior research, in the ecological sciences and in many other fields
[Bishop, Fienberg and Holland (\citeyear{Bishop1995}), Bunge and Fitzpatrick (\citeyear{Bunge1993}),
Chao et al. (\citeyear{Chao2000}), Hay and Smit (\citeyear{Hay2003}), Pledger (\citeyear{Pledger2000,Pledger2005}),
Roberts and Brewer (\citeyear{Roberts2006}), Wilson and Collins (\citeyear{Wilson1992})].
A~prominent problem in public health is the completeness of a disease
registry [Van Hest et al. (\citeyear{Hest2007})], while an interesting application
of capture--recapture techniques in the veterinary sciences is the
estimation of hidden scrapie in Great Britain [B\"{o}hning and Del Rio
Vilas (\citeyear{Boening20082})].
In software engineering [Wohlin, Runeson, and Brantestam (\citeyear{Wohlin1995})] we are
interested in finding the number of errors hidden in software
components. In criminology the number of people with illegal behavior
is of high interest [Van der Heijden, Cruyff, and Houwelingen (\citeyear{Heijden2003})],
and in ecology we wish to estimate the number of rare species of
organisms [Chao et al. (\citeyear{Chao2000})]. All of these situations fall under
the following setting. We assume that there are $N$ units in the
population, which is closed (no birth, death or migration), and that
there is an endogenous mechanism such as a~register, a~diagnostic
device, a set of reviewers, or a trapping system, which identifies $n$
distinct units from the population. A given unit may be identified
exactly once, or it may be observed twice, three times, or more. We
denote the number of units observed $i$ times by $f_i$, so that
$n=f_1+f_2+f_3+\cdots$; the number of unobserved or missing units is
$f_0$, so $N = f_0 + n$. The objective is to find an estimate (or
rather a prediction) $\hat{f}_0$ for $f_0$, and hence an estimate
~$\hat{N}$ of~$N$.\looseness=-1

To illustrate, we first introduce several examples from different
domains; these are analyzed in the following sections:
\begin{enumerate}
\item\textit{Methamphetamine use in Thailand.} Surveillance data on drug
abuse are available for 61 health treatment centers in the Bangkok
metropolitan region from the Office of the Narcotics Control Board
(ONCB). Using this data, it was possible to reconstruct the counts of
treatment episodes for each patient in the last quarter of 2001. Table
\ref{tab:methtable} presents the number of methamphetamine users for
each count of
treatment episodes [B\"{o}hning et al. (\citeyear{Boehning2004})]; the maximum observed
frequency was 10. Here we are interested in estimating the number of hidden
methamphetamine users.
%
%
\begin{table}
\caption{Methamphetamine data---frequency distribution of treatment
episodes per drug user}\label{tab:methtable}
\begin{tabular*}{\textwidth}{@{\extracolsep{\fill}}lccccccccc@{\hspace*{3pt}}c@{}}
\hline
$\bolds{f_{1}}$ & $\bolds{f_{2}}$ & $\bolds{f_{3}}$ & $\bolds{f_{4}}$ & $\bolds{f_{5}}$ & $\bolds{f_{6}}$ & $\bolds{f_{7}}$ &
$\bolds{f_{8}}$ & $\bolds{f_{9}}$ & $\bolds{f_{10}}$ & $\bolds{n}$ \\
\hline
3114 & 163 & 23 & 20 & 9 & 3& 3 & 3 & 4 & 3 & 3345 \\
\hline
\end{tabular*}
\end{table}
\item\textit{Screening for colorectal polyps.} In 1990, the Arizona
Cancer Center initiated a multicenter trial to determine whether wheat
bran fiber can prevent the recurrence of colorectal adenomatous polyps
[Alberts et al. (\citeyear{Alberts2000}), Hsu (\citeyear{Hsu2007})]. Subjects with previous history
of colorectal adenomatous polyps were recruited and randomly assigned
to one of two treatment groups, low fiber and high fiber. The
researchers noted that adenomatous polyp data are often subject to
unobservable measurement error due to misclassification at colonoscopy.
It can be assumed that patients with a positive polyp count were
diagnosed correctly, whereas it is unclear how many persons with
zero-count of polyps were false-negatively diagnosed. Thus, we approach
the data as if zero-counts were not observed, and we try to estimate
the undercount from the nonzero frequencies. Table \ref{tab:polyptable}
shows the polyp frequency data for the two different treatment groups;
the (overall) maximum frequency is 77. The number of subjects with an
\textit{observed} number of adenomas equal to 0 is $285$ for the Low Fiber
treatment and $381$ for High Fiber treatment respectively; we regard
this as an undercount and seek to estimate the true unobserved
frequencies $f_0$.
%
%
\begin{table}[b]
\tabcolsep=0pt
\caption{Polyps data---frequency distribution of recurrent
adenomatous polyps per patient, by treatment group}\label{tab:polyptable}
\begin{tabular*}{\textwidth}{@{\extracolsep{\fill}}lccccccccccccc@{\hspace*{3pt}}c@{}}
\hline
& $\bolds{(f_ {0})}$ & $\bolds{f_{1}}$& $\bolds{f_{2}}$ & $\bolds{f_{3}}$ & $\bolds{f_{4}}$ & $\bolds{f_{5}}$ & $\bolds{f_{6}}$
& $\bolds{f_{7}}$ & $\bolds{f_{8}}$ & $\bolds{f_{9}}$ &$\bolds{f_{10}}$&$\bolds{f_{11}}$&$\bolds{\cdots}$ & \\
\hline
Low & (285) & 145 & 66 & 39 & 17 & \phantom{0}8 & 8 & 7 & 3 & 1 & 0 & 3 &
$\cdots$&\\
High& (381) & 144 & 61 & 55 & 37 & 17 & 5 & 4 & 6 & 5 & 1 & 1 &
$\cdots$&\\[6pt]
& $\bolds{f_ {22}}$ & $\bolds{\cdots}$ & $\bolds{f_{28}}$ & $\bolds{\cdots}$ & $\bolds{f_{31}}$ & $\bolds{\cdots}$ &
$\bolds{f_{44}}$ & $\bolds{\cdots}$ & $\bolds{f_{57}}$ & $\bolds{\cdots}$ & $\bolds{f_{70}}$ & $\bolds{\cdots}$ &
$\bolds{f_{77}}$ & $\bolds{n}$ \\
\hline
Low & \phantom{00}1 & $\cdots$ & \phantom{0}1 & $\cdots$ & \phantom{0}0 & $\cdots$ & 0 & $\cdots
$ &
0 & $\cdots$ & 0 & $\cdots$ & 0 & 299\\
High & \phantom{00}0 & $\cdots$ & \phantom{0}0 & $\cdots$ & \phantom{0}1 & $\cdots$ & 1 & $\cdots$
& 1 & $\cdots$ & 1 & $\cdots$ & 1 & 341\\
\hline
\end{tabular*}
\end{table}
\item\textit{Scrapie in Great Britain.}
Sheep are kept in holdings in Great Britain and the occurrence of
scrapie in the population of holdings is monitored by the Compulsory
Scrapie Flocks Scheme [B\"{o}hning and Del Rio Vilas (\citeyear{Boening20082})]. This was
established in 2004 and summarizes three surveillance sources. Table
\ref{tab:scrapietable} presents the frequency distribution of the \textit{
scrapie count within each holding} for the year 2005. Here interest is
estimating $f_0$, the frequency of holdings with unobserved or
unreported scrapie. The maximum frequency in the data is 8.
%
%
\begin{table}
\tablewidth=250pt
\caption{Scrapie data---frequency distribution of the
scrapie count within each holding for Great Britain in 2005}\label{tab:scrapietable}
\begin{tabular*}{250pt}{@{\extracolsep{\fill}}lccccccc@{\hspace*{3pt}}c@{}}
\hline
$\bolds{f_{1}}$ & $\bolds{f_{2}}$ & $\bolds{f_{3}}$ & $\bolds{f_{4}}$ & $\bolds{f_{5}}$ & $\bolds{f_{6}}$ & $\bolds{f_{7}}$ &
$\bolds{f_{8}}$ & $\bolds{n}$ \\
\hline
84 & 15 & 7 & 5 & 2 & 1 & 2 & 2 & 118\\
\hline
\end{tabular*}\vspace*{-5pt}
\end{table}
\item\textit{Malayan butterfly data.} This data set derives from a large
collection of Malayan butterflies collected by A. S. Corbet in 1942
[Fisher, Corbet and Williams (\citeyear{Fisher1943})]. There were 9031 individual
butterflies classified to $n=620$ species. Out of these 620 different
species, 118 were observed exactly once, 74 twice, 44 three times and
so forth. This ``abundance'' data is shown in Table \ref
{tab:butterfliestable}. Fisher, Corbet and Williams (\citeyear{Fisher1943}) reported
exact counts only up to $f_{24}$, stating that there were a total of
119 species with sample abundances (counts) greater than 24. Here the
interest is in estimating the total number of species $N$.
%
%
\begin{table}[b]\vspace*{-4pt}
\caption{Butterfly data---frequency distribution of
butterfly species collected in Malaya}\label{tab:butterfliestable}
\begin{tabular*}{\textwidth}{@{\extracolsep{\fill}}lcccccccccccc@{\hspace*{10pt}}c@{}}
\hline
$\bolds{f_{1}}$ & $\bolds{f_{2}}$ & $\bolds{f_{3}}$ & $\bolds{f_{4}}$ & $\bolds{f_{5}}$ & $\bolds{f_{6}}$ & $\bolds{f_{7}}$ &
$\bolds{f_{8}}$ & $\bolds{f_9}$ & $\bolds{f_{10}}$ & $\bolds{f_{11}}$& $\bolds{f_{12}}$ & &\\
\hline
118 & 74 & 44 & 24 & 29 & 22 & 20 & 19 & 20 & 15 & 12 & 14 & & \\ [6pt]
$\bolds{f_{13}}$
&$\bolds{f_{14}}$&$\bolds{f_{15}}$&$\bolds{f_{16}}$&$\bolds{f_{17}}$&$\bolds{f_{18}}$&$\bolds{f_{19}}$&$\bolds{f_{20}}$&$\bolds{f_{21}}$
&$\bolds{f_{22}}$&$\bolds{f_{23}}$&$\bolds{f_{24}}$
& $\bolds{f_{>24}}$ & $\bolds{n}$ \\
6 & 12 & 6 & 9 & 9 & 6 & 10 & 10 & 11 & 5 & 3 & 3 & 119 & 620 \\
\hline
\end{tabular*}
\end{table}
\item\textit{Microbial diversity in the Gotland Deep.} The data on
microbial diversity shown in Table \ref{tab:microbialtable} stem from a
recent work by Stock et al. (\citeyear{Stock2009}). Microbial ecologists are
interested in estimating the number of species $N$ in particular
environments. Unlike butterflies, microbial species membership is not
clear from visual inspection, so individuals are defined to be members
of the same species (or more general taxonomic group) if their DNA
sequences (derived from a certain gene) are identical up to some given
percentage, 95\% in this case. Here the study concerned protistan
diversity in the Gotland Deep, a basin in the central Baltic Sea. The
sample was collected in May 2005. The maximum observed frequency was~53.
\end{enumerate}

The classical approach to estimation of $N$ is to assume that each
population unit enters the sample independently with probability $p$
(dealing with heterogeneous capture probabilities by modeling and
averaging). Given $p$, the unbiased Horvitz--Thompson estimator of $N$
is $n/p$, and the maximum likelihood estimator is its integer part
$\lfloor n/p \rfloor$. One then estimates $p$ using any of several
methods, and the final estimate of $N$ is $n/\hat{p}$ or $\lfloor n/\hat
{p} \rfloor$ [Lindsay and Roeder (\citeyear{Lindsay1987}), B\"{o}hning et al. (\citeyear{Boehning2005}),
B\"{o}hning and van der Heijden (\citeyear{Boehning2009}), Wilson and Collins (\citeyear{Wilson1992}),
Bunge and Barger (\citeyear{Bunge2008}), Chao (\citeyear{Chao1987,Chao1989}), Zelterman (\citeyear{Zelterman1988})].

\begin{figure}

\includegraphics{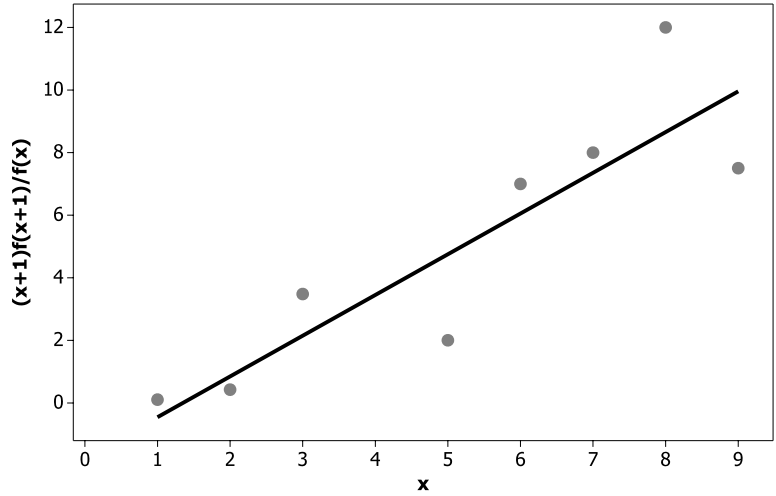}
\vspace*{-6pt}
\caption{Scatterplot with regression line of
$(x+1)f_{(x+1)}/f_x$ vs. $x$ for the Bangkok methamphetamine drug user data.}
\label{fig:figure1}
\vspace*{-6pt}
\end{figure}

Here we take a new approach: we consider 
\textit{ratios of successive frequency counts}, namely,
\[
\hat{r}(x):= \frac{(x+1)f_{x+1}}{f_x}.
\]
Often $\hat{r}(x)$ appears as a roughly linear function of $x$, which
leads us to apply linear regression to the scatterplot of $(x,\hat
{r}(x))$; we then project the regression function downward to the left,
to zero, which yields $\hat{f}_0$ and hence~$\hat{N}$. Figure~\ref
{fig:figure1} shows the \textit{ratio plot} of $(x,\hat{r}(x))$ for the
methamphetamine data; there is clear evidence for a linear trend.
Projecting the line to the left, we obtain $\hat{f}_0 = 57\mbox{,}788$ and,
hence, $\hat{N}=61\mbox{,}133$.
%
%
\begin{table}
\caption{Protistan diversity in the Gotland
Deep---frequency counts of observed species}\label{tab:microbialtable}
\vspace*{-5pt}
\begin{tabular*}{\textwidth}{@{\extracolsep{\fill}}lcccccccc@{\hspace*{10pt}}c@{}}
\hline
$\bolds{f_{1}}$ & $\bolds{f_{2}}$ & $\bolds{f_{3}}$ & $\bolds{f_{4}}$ & $\bolds{f_{6}}$
& $\bolds{f_{8}}$ & $\bolds{f_{9}}$ & $\bolds{f_{10}}$ & $\bolds{f_{11}}$ & \\
\hline
48 & 9 & 6 & 2 & 2 & 2 & 1 & 2 & 1 & \\ [6pt]
$\bolds{f_{12}}$ & $\bolds{f_{13}}$ & $\bolds{f_{16}}$ & $\bolds{f_{17}}$ & $\bolds{f_{18}}$ & $\bolds{f_{20}}$ &
$\bolds{f_{29}}$ & $\bolds{f_{42}}$ & $\bolds{f_{53}}$ & $\bolds{n}$ \\
\hline
\phantom{0}1 & 1 & 2 & 1 & 1 & 1 & 1 & 1 & 1 & 84 \\
\hline
\end{tabular*}
\vspace*{-5pt}
\end{table}
%

\begin{figure}[b]
\vspace*{-6pt}
\includegraphics{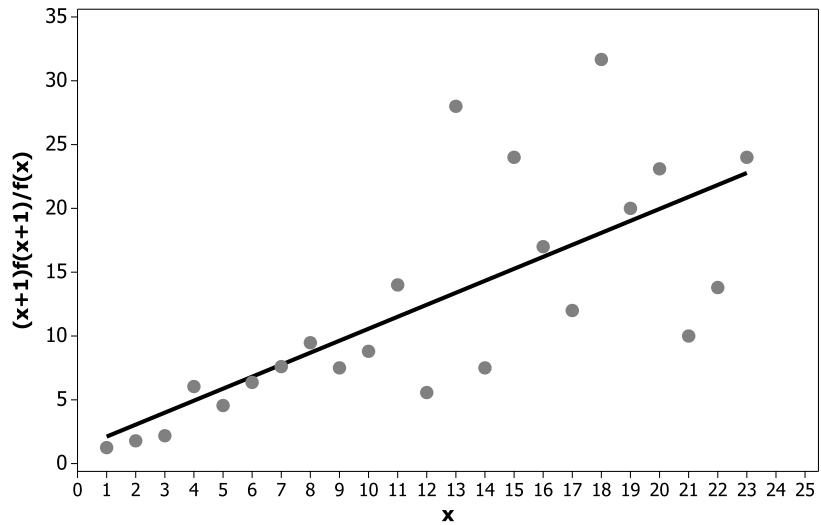}
\vspace*{-6pt}
\caption{Scatterplot with regression line of
$(x+1)f_{(x+1)}/f_x$ vs. $x$ for the butterfly data.}
\label{fig:figure2}
\end{figure}

Figure \ref{fig:figure2} shows the ratio plot for the butterfly data;
again there is a clear linear trend and here we also observe increasing
variance in the points as $x$ increases, which we will deal with via
weighted least squares. In this case we find $\hat{f}_0 = 126$ and $\hat
{N} = 746$.
%
%

This simple and powerful method applies exactly when the frequency
counts emanate from the Katz family of distributions, namely, the
binomial, Poisson and gamma-mixed Poisson or negative binomial, and it
applies approximately to extensions of the Katz family and to general
Poisson mixtures. It can be implemented using any statistical software
package that performs weighted least squares regression, and it is
superior to existing methods for the negative binomial model (including
maximum likelihood) in several ways. In addition, it substantially
mitigates the effect of truncating large counts (recaptures or
replicates), which is an issue with almost every existing method,
parametric or nonparametric. In Section~\ref{sec2} we discuss the method and its
scope of applicability; in Section \ref{ls} we describe weighting schemes; in
Section \ref{model} we look at goodness of fit of the linear model; and in
Section \ref{sim} we compare our method with existing techniques, analyze the
five data sets, and discuss the implications of our findings.
The \hyperref[app]{Appendix} covers aspects of the approximation used for reaching the
linear model as well as a comparative simulation study, a discussion of
standard error approximations, and an assessment of the effect of
deleting large ``outlying'' frequencies.

\section{Linear regression and the Katz distributions}\label{sec2}
Let $p_0, p_1, p_2, \ldots$ denote a probability distribution on the
nonnegative integers. The condition
%
%
\begin{equation}
\label{katz}
\frac{(x+1)p_{x+1}}{p_x} = \gamma+ \delta x, \qquad x = 0, 1,
2, \ldots,
\end{equation}
where $\gamma$ and $\delta$ are real constants, characterizes the \textit{
Katz family of distributions} [Johnson, Kemp and Kotz (\citeyear{Johnson2005})].
To yield a valid probability\vadjust{\goodbreak} distribution, it is necessary that $\gamma
> 0$ and $\delta< 1$. If $\delta< 0$, $p_x$ is the binomial
distribution; if $\delta= 0$, $p_x$ is the Poisson; and if $\delta\in
(0,1)$, $p_x$ is the negative binomial. These distributions arise
naturally as models for population size estimation.
\begin{itemize}
\item Suppose that a given population unit may be observed on each of
$k$ ``trapping occasions.'' Assume further that the trapping or capture
probability, say, $r$, is the same on each occasion and that captures
are independent across occasions, and also that the capture probability
is the same (homogeneous) for all units, and that units are captured
independently of each other. If $m_i$ denotes the number of captures of
the $i$th unit, then $m_1,\ldots,m_N$ are i.i.d. binomial ($k,r$)
random variables. This simple model is rarely realistic, but it can
provide a lower bound for the population size, since the homogeneity
assumption leads to downwardly biased estimation in the presence of
heterogeneity. This is formally proved in B\"{o}hning and
Sch\"{o}n (\citeyear{Boehning20051}) for maximum likelihood estimation. In this case the
frequency count data $f_1,f_2,\ldots$ summarizes the nonzero values of
$m_1,\ldots,m_N$.
\item Now suppose that population unit $i$ appears a random number of
times $m_i$ in the sample, but now $m_1,\ldots,m_N$ are i.i.d. Poisson
random variables with (homogeneous) mean $\lambda$. This model arises
naturally in \textit{species abundance sampling} where each species
contributes some number of representatives to the sample; it also
appears as an approximation to the binomial model with $\lambda\approx
kr$, for large $k$ and small $r$. Again the homogeneity makes this
model mainly useful for lower-bound benchmarking.
\item Assume now that the foregoing Poisson model holds, but with the
modification that the mean number of appearances of unit $i$ is $\lambda
_i$, and that $\lambda_1,\ldots,\lambda_N$ are i.i.d.
gamma-distributed random variables. Then the distribution of $m_i$ is
(unconditionally) gamma-mixed Poisson, that is, negative binomial. This
is not the simplest possible model with heterogeneous capture rates,
but it may be the oldest, appearing in Fisher, Corbet and Williams
(\citeyear{Fisher1943}), the source of our butterfly data. (Note that it includes the
geometric, since the exponential is a special case of the gamma.) The
negative binomial distribution is widely applicable as a model for the
frequency counts, when the data is not too highly skewed (left or
right); however, it is surprisingly difficult to fit by, for example,
maximum likelihood, or by other existing procedures such as the
Chao--Bunge estimator (see discussion below). We show below that, when
implemented by our weighted least squares regression procedure, the
negative binomial model becomes practical and useful for estimating $N$
in a variety of situations.
\end{itemize}

We make two further comments on distribution theory. First, it may be
readily shown using the Cauchy--Schwarz inequality that the ratio on
the left-hand side of (\ref{katz}) is nondecreasing for \textit{any}
mixed-Poisson distribution. This means that the linear relation, and
hence our weighted linear regression procedure below, can be regarded
as a first-order linear approximation for any Poisson mixture (not just
gamma), thus justifying a degree of robustness of our method across a
wide range of heterogeneity models. Second, there are extended versions
of relation (\ref{katz}) which give rise to distributional extensions
of the Katz family that need not be mixed-Poisson [Johnson, Kemp and
Kotz (\citeyear{Johnson2005})]. Such extensions may be parameterized and we conjecture
that our method below will be robust to small perturbations along these
parameters.\looseness=1

Condition (\ref{katz}) suggests linear regression of the left-hand side
upon the right, in some form. Observe that the natural estimate of
$p_x$ would be $\hat{p}_x(N):=f_x/N$, if $N$ were known. But
\[
\frac{(x+1)\hat{p}_{x+1}(N)}{\hat{p}_x(N)} = \frac
{(x+1)f_{x+1}/N}{f_x/N} = \frac{(x+1)f_{x+1}}{f_x} = \hat{r}(x),
\]
so we can fit a linear regression of $\hat{r}(x)$ on $x$ without
knowing $N$. We can then obtain an estimate of $f_0$ by setting $x=0$
so that $\hat{r}(0) = 1f_1/\hat{f}_0 = \hat{\gamma}$, and, hence, $\hat
{f}_0 = f_1/\hat{\gamma}$.
In practice, however, we prefer to fit the response on a logarithmic
scale, which is approximately linear near the origin and avoids
negative fitted values. Thus, our basic equation becomes
\[
\log\biggl(\frac{(x+1)p_{x+1}}{p_x}\biggr) = \gamma+ \delta x ,
\]
and we fit the model
%
%
\begin{equation}
\log\biggl(\frac{(x+1)f_{x+1}}{f_x}\biggr) = \gamma+ \delta x +
\varepsilon_x .
\label{model_1}
\end{equation}
We consider this in terms of linear regression in the next section. The
estimate of~$f_0$ is then $\hat{f}_0 = f_1 e^{-\hat{\gamma}}$.

In particular, consider the gamma-mixed Poisson or negative binomial
model for the count data. Let the negative binomial be parameterized as
\[
p(x) = \frac{\Gamma(x+k)}{\Gamma(x+1)\Gamma(k)}p^k(1-p)^x,
\]
where $k>0$ and $p\in(0,1)$. Similar to other areas such as Poisson
regression, we need to apply a suitable transformation to avoid
negative values for the ratios which would lead to negative estimates
for $f_0$. The log-transformation is appropriate, although others are
also possible. Transforming both sides, we obtain
\[
\log\{(x+1)p(x+1)/p(x)\} = \log(x+k)+\log(1-p) ,
\]
but now the right-hand side is nonlinear in $k$. However, taking the
first-order Taylor expansion of $\log(k+x)$ around $k$, we achieve
\[
\log(k+x) \approx\log(k) + \frac{1}{k}x,
\]
so that we have $\log(x+k)+\log(1-p) \approx\log(1-p) + \log(k) + x/k$. Note that this approximation is exact for $x=0$ (the
point where we predict) and good for $x=1$ (corresponding to the
informative ``singleton'' frequency count). In the \hyperref[app]{Appendix} we discuss
this approximation further, as well as alternatives. With reference to
model (\ref{model_1}), we have $\gamma= \log(1-p) + \log(k)$ and
$\delta= 1/k$. We focus on this model in the discussion below.

Note also that due to the simple structure of the estimator $\hat f_0 =
f_1 \exp(-\hat\gamma)$, we can use conditioning [B\"{o}hning (\citeyear{Boehning2007})]
in combination with the $\delta$-method to give an approximate
expression for the variance of $\hat f_0$ as
\[
\operatorname{Var}(\hat f_0) \approx\exp(-\hat\gamma)^2f_1[\operatorname{Var}(\hat
\gamma)f_1+1],
\]
where $\operatorname{Var}(\hat\gamma)$ is the variance of the intercept
estimator in the regression model. An approximation to the variance of
$\hat N= \hat f_0 + n$ is then [using the same technique and estimating
$\operatorname{Var}(n) = N(1-p_0)p_0$ by $n\hat f_0/\hat N$]
%
%
\begin{equation}
\operatorname{Var}(\hat N) \approx n\frac{\hat f_0}{\hat N}+\exp(-\hat\gamma
)^2f_1[\operatorname{Var}(\hat\gamma)f_1+1].
\label{variance}
\end{equation}
Standard errors are obtained by plugging in estimates for $\operatorname{Var}(\hat{\gamma})$ and taking the (overall) square root. These
expressions may be imprecise for small sample sizes ($<$100) and in
such cases the bootstrap might be preferable. We provide a simulation
study on this aspect in the \hyperref[app]{Appendix}.

\section{Heteroscedasticity and weighted least squares}\label{ls}

Model (\ref{model_1}) does not satisfy the classical linear regression
assumptions. In the first place, the response is discrete (although
log-transformed), so we might consider a generalized linear model such
as Poisson or even negative binomial regression. However, this is
inadvisable since an appropriate formulation as a generalized linear
model leads to an autoregressive equation involving $\log f_x$ as an
additional offset term in the linear predictor. These kinds of models
experience difficulties in terms of the definition of the likelihood as
well as in carrying out inference. Actually, residuals derived from
model (\ref{model_1}) typically show reasonable conformity with normal
probability plots when the linear model fits well (see Section \ref{model}
regarding goodness of fit). The issues of dependence and
heteroscedasticity are more important, and we address these by using
weighted least squares. We take
\[
\pmatrix{\hat\gamma\cr
\hat\delta
}
=(\mathbf{X}^T \mathbf{W}\mathbf{X})^{-1}\mathbf{X}^T \mathbf{W}
\mathbf{Y},
\]
where
\[
\mathbf{Y}=
\pmatrix{
\log\bigl(\frac{2 f_{2}}{f_{1}}\bigr) \vspace*{2pt}\cr
\log\bigl(\frac{3
f_{3}}{f_{2}}\bigr) \vspace*{2pt}\cr
\vdots\vspace*{2pt}\cr\log\bigl(\frac{m
f_{m}}{f_{m-1}}\bigr)},\qquad
\mathbf{X}=
\pmatrix{
1 &1
\cr
1 &2
\cr
\vdots& \vdots
\cr1 &m-1},
\]
and $m$ is the maximum frequency used in the estimator (see Section \ref{model}
below regarding truncation of large frequencies). To reduce MSE, we
wish to take $\mathbf{W} \approx(\operatorname{cov}\!\mathbf{(Y)})^{-1}$. To
find $\operatorname{cov}\!\mathbf{(Y)}$, assume that the distribution of the cell counts
$f_{1},\ldots,f_{m}$ is multinomial
with cell probabilities $\pi=(\pi_{1},\ldots,\pi_{m})^T$. Then it is
well known that
$\mathbf{f}=(f_1,\ldots,f_m)^T $
has covariance matrix $\Sigma=n[\Lambda(\mathbf{\pi})-\mathbf{\pi}
\mathbf{\pi}^T]$,
where $\Lambda(\mathbf{\pi})$ is a diagonal matrix with elements
$\mathbf{\pi}$ on the diagonal, and $n=f_1+\cdots+f_m$. Writing
\[
\Sigma=n[\Lambda(\mathbf{\pi})-\mathbf{\pi} \mathbf{\pi}^T]=\Lambda
(n\mathbf{\pi})-\frac{1}{n}n\mathbf{\pi} n\mathbf{\pi}^T,
\]
we see that $\Sigma$ can be estimated as
\[
\hat\Sigma=\Lambda(\mathbf{f})- \frac{1}{n} \mathbf{f}\ \mathbf{f}^T .
\]
An application of the multivariate delta-method then shows that an
estimate of $\operatorname{cov}(\mathbf{Y})$ is
%
%
\begin{eqnarray}\label{full}
\qquad &&\nabla_{\mathbf{f}}(\mathbf{Y}(\mathbf{f}))\hat{\Sigma}
(\nabla_{\mathbf{f}}^T(\mathbf{Y}(\mathbf{f)}))
\nonumber
\\[-8pt]
\\[-8pt]
\nonumber
&&\qquad=
\left[\matrix{
\frac{1}{f_{1}}+\frac{1}{f_{2}} & \frac{-1}{f_{2}} & 0 & \ldots& 0
&\ldots& 0\cr
\frac{-1}{f_{2}}& \frac{1}{f_{2}}+\frac{1}{f_{3}}& \frac{-1}{f_{3}}
&0& & \ldots& 0\cr
0 & & \ddots& & & & \cr
\vdots& & & \ddots& & & \cr
0 & \ldots\ 0 & \frac{-1}{f_{i}} & \frac{1}{f_{i}}+\frac{1}{f_{i+1}} &
\frac{-1}{f_{i+1}}&0\ \ldots& 0 \cr
\vdots& & & & \ddots&& \cr
0& & && 0& \frac{-1}{f_{m-1}} &
\frac{1}{f_{m-1}}+\frac{1}{f_{m}}}\right]\!.
\vspace*{3pt}
\end{eqnarray}
Note that this requires that only nonzero frequencies be used in the
estimate.\looseness=1\vfill\eject

The tridiagonal matrix (\ref{full}) has a special structure, and
Meurant (\citeyear{meurant1992}) gives an analytical formula for its inverse. In
addition, a calculation based on the representation in Meurant's
Theorem 2.3 indicates that it may be possible to drop the off-diagonal
terms in $\operatorname{cov}(\mathbf{Y})$ with little loss of numerical
precision for our purposes. This corresponds to our intuition that
covariances between adjacent log-ratios may not play a large role in
reducing MSE. Let
%
%
\begin{equation}
\Lambda(\mathbf{f})=
\left[\matrix{
\frac{1}{f_{1}}+\frac{1}{f_{2}} & 0 & 0 & \ldots& 0 & 0\cr
0 & \frac{1}{f_{2}}+\frac{1}{f_{3}}& 0 & \ldots& 0 & 0\cr
\vdots& & \ddots& & & \cr
\vdots& & & \ddots& & \cr
0 & 0 & 0 & \frac{1}{f_{i}}+\frac{1}{f_{i+1}} & 0 & 0 \cr
\vdots& & & & \ddots& \cr
\ldots& & & & 0 & \frac{1}{f_{m-1}}+\frac{1}{f_{m}}}\right]
\label{dia}
\end{equation}
be the diagonal part of (\ref{full}); we then suggest using (\ref{dia})
in our weighted regression model. This is computationally simpler,
especially when dealing with a high number of recaptures. A small
simulation study confirms the precision of this simplification, at
least within the domain of the simulation. We computed the bias of $\hat
{N}$ using the weighted regression model under three scenarios: with
weights according to (\ref{full}), according to (\ref{dia}) and
according to $\mathbf{W}= I_{m-1}$ [the $(m-1)$-dimensional identity
matrix, i.e., unweighted]. Frequency data were drawn from a negative
binomial distribution with parameters $p=0.8$ and $k=7$, and replicated
$1000$ times. Table \ref{table7} shows results for $N=100$ and $N=1000$.
%
%
\begin{table}[b]
\tablewidth=250pt
\caption{The effect of different weight
matrices according to (\protect\ref{full}), (\protect\ref{dia}) and $\mathbf
{W}=I_{m-1}$ for frequency data from the Negative Binomial
distribution with parameters $k=7$, $p=0.8$}\label{table7}
\begin{tabular*}{250pt}{@{\extracolsep{\fill}}ld{2.2}d{2.2}d{2.2}@{}}
\hline
\multicolumn{1}{@{}l}{$\bolds{N}$}& \multicolumn{1}{c}{\textbf{(\ref{full})}}& \multicolumn{1}{c}{\textbf{(\ref{dia})}} & \multicolumn{1}{c@{}}{\textbf{Unweighted}} \\
\hline
\multicolumn{4}{@{}l}{Bias of $\hat N$ }\\
\phantom{0}100 & 3.05 & 3.40 & 8.81 \\
1000 & 2.70& 0.36 & 45.86 \\[6pt]
\multicolumn{4}{@{}l}{Standard error of $\hat N$ }\\
\phantom{0}100& 10.48 & 11.73 & 13.79 \\
1000 & 29.12 & 32.04 & 56.87 \\
\hline
\end{tabular*}\vspace*{-4pt}
\end{table}
It is clear that weighting is important in fitting the model: the
unweighted regression model leads to potentially heavily biased
estimators of the population size, whereas the effect of ignoring the
covariance between $\log(xf_x/f_{x-1})$ and $\log((x+1)f_{x+1}/f_{x})$ is negligible. Finally, we note that
weighted least squares can introduce numerical problems, especially in
sparse-data situations [Bj\"{o}rck (\citeyear{Bjorck1996}), Chapters 4 and 6]; however,
our design matrix has only rank 2 and our maximum frequency $m$ is
typically not too large, so we have not yet encountered such problems
here. This is a topic for future research in this context.

\section{Model assessment and goodness of fit}\label{model}

The ratio plot shown in Figure \ref{fig:figure1} is our main graphical tool for looking
at goodness of fit of the linear regression model, and having fit the
model, the standard diagnostic plots of residuals are also available.
We also require a quantitative assessment of overall fit: $R^2$ could
be used based on the response $\log(x+1)f_{x+1}/{f_x}$, but in this
setting it seems more appropriate to work on the original frequency of
counts scale. In addition, we are looking for a measure which allows
analysis of residuals. We therefore compare the observed frequencies
with the estimated frequencies from the model, using the $\chi
^{2}$-statistic as a goodness-of-fit measure [Agresti (\citeyear{Agresti2002})].
The estimated frequencies based on the regression model are
\[
\hat{y}_x =\log\widehat{\frac{(x+1)f_{x+1}}{f_{x}}} =\hat{\gamma}+\hat
{\delta}x ,\vadjust{\goodbreak}
\]
$x=1,2,\ldots,m$, or, equivalently,
\[
\widehat{\frac{(x+1)f_{x+1}}{f_{x}}}=\exp(\hat{y}_x),
\]
where $m$ is the ``truncation point'' or maximum frequency used in the
analysis (we return to this issue below).
In general, the estimated ratios of frequencies $\widehat{f_{x+1}/
f_{x}}$ need not uniquely determine $\hat f_{x+1}$ and $\hat
f_{x}$, but in this case they do since $\hat{f}_0 = f_1/\exp(\hat
{\gamma}) =f_1/( \widehat{f_{1}/\hat f_{0}})$. This also shows that
$\hat f_1=f_1$, since $\hat{f}_0 = f_1/\exp(\hat{\gamma}) = f_1 /\exp
(\hat{y}_0)$, and, hence, $\hat{f}_1 =
\hat{f}_0\exp(\hat{y}_0) = f_1$. Now, with $\hat f_1$ given the
equation $2\hat f_2/\hat f_1 = \widehat{2f_{2}/ f_{1}}$ determines
$\hat f_2$ uniquely,
leading to the recursive relation $\hat{f}_{x+1}=\hat{f}_x\exp(\hat
{y}_x)/(x+1)$, $x=1,2,\ldots,m-1$.
We then define our $\chi^2$ statistic as
\[
\chi^2 = \sum_{x=1}^m \frac{(f_x-\hat f_x)^2}{\hat f_x}
\]
and simulations support that this has a $\chi^2$ distribution with
$m-2$ degrees of freedom if the regression model $y_x = \gamma+ \delta
x$ is correct. Note that we have~$m$ unconstrained frequencies, since
$n= \sum_{x=1}^m f_x$ is random, and we lose $2$ degrees of freedom due
to estimating the intercept and slope parameters. Note also that the
estimate of the intercept parameter fixes $\hat f_1 =f_1$, so that the
degrees of freedom are indeed only reduced by 2.
This approach has the benefit of gaining one degree of freedom when
compared to a goodness-of-fit
measure based solely on the regression model which works with the $m-1$
values $\hat y_x$, $x=1,\ldots,m-1$.

This argument is conditional upon fixing the value of $m$, and, indeed,
all known procedures for population size estimation truncate large
``outlier'' frequencies in some way. To illustrate, we return to the
classical maximum likelihood (ML) approach. Bunge and Barger (\citeyear{Bunge2008})
describe a procedure which fits the desired distribution (here, the
negative binomial) to the (nonzero) frequency count data by ML; the
estimate of $N$ is then based upon the estimated parameter values of
the distribution. Typically, parametric distributions can only be made
to fit the data up to some truncation point~$m$, beyond which the fit,
as assessed by the classical Pearson $\chi^2$ test, falls off
considerably; consequently, only frequencies up to $m$ are used to
obtain the estimate of $N$, and the number of units with frequencies
greater than $m$ is added to the estimate \textit{ex post facto}. Bunge
and Barger (\citeyear{Bunge2008}) propose a~goodness-of-fit criterion for selecting
$m$, while the coverage-based nonparametric methods of Chao and
co-authors fix $m$ heuristically at 10 [see Chao and Bunge (\citeyear{ChaoBunge2002})]. Our
weighted linear regression approach also has the potential for loss of
fit as $m$ increases, depending on the realized structure of the data,
and again we can fix $m$ prior to the analysis, and collapse all
frequencies greater than this threshold to one value. Sensitivity of
the various methods to the choice of $m$ is a complex topic [Bunge and
Barger (\citeyear{Bunge2008}) compute all estimates at all possible values of $m$];
however, our data analyses below show that the weighted linear
regression model is considerably less sensitive to $m$ than its chief
competitors in the negative binomial case, namely, ML and the
Chao--Bunge estimator.

Finally, we note that in the ML approach, if the negative binomial fit
is less than ideal (although perhaps still acceptable), numerical
maximum likelihood algorithms often do not converge, or converge to the
edges of the parameter space, which in turn distorts the apparent fit.
The regression-based method described here offers a more robust
approach to parameter estimation, and appears not to be prone to the
numerical problems which arise for maximum likelihood estimation under
the negative binomial model. In fact, the negative binomial parameter
estimates $(\hat{p},\hat{k})$ derived from the regression model and
could be used as starting values for a numerical search for the ML
estimates. This is a topic for further research.

\section{Alternative estimators, data analyses and discussion} \label{sim}
\subsection{Alternative estimators}
We first consider certain other options for the negative binomial model.
\begin{itemize}
\item\textit{Maximum likelihood.} This approach is well studied and has
a long history [see Bunge and Barger (\citeyear{Bunge2008})], but as noted above, good
numerical solutions for the model parameters $(p,k)$ seem to be
remarkably difficult to obtain, even using reasonably sophisticated
search algorithms with high-precision settings. In our experience we
get good numerical convergence only when the frequency data is smooth
and fits the negative binomial well, or the right-hand tail is fairly
severely truncated. The latter issue causes the additional
computational burden of investigating a diversity of truncation points,
each involving numerical optimization. Nonetheless, we can obtain ML
results for the negative binomial in some cases. The ML estimator $\hat
{N}_{\mathrm{ML}}$ is consistent for $N$ given that the model is correct.
\item\textit{Chao--Bunge.} Let $\tau$ denote the probability of observing
a unit at least twice, that is, $\tau= 1-p_0-p_1$. Chao and Bunge
(\citeyear{ChaoBunge2002}) developed a nonparametric estimator $\hat{\tau}$ for $\tau$, and
on this basis proposed the estimator
\[
\hat{N}_{\mathrm{CB}}:= \sum_{j=2}^m \frac{f_j}{\hat{\tau}}
\]
for $N$. They showed that $\hat{N}_{\mathrm{CB}}$ is consistent for $N$ under
the negative binomial model. However, in applied data analysis $\hat
{\tau}$ may be very small or even negative, leading to very large or
negative values of $\hat{N}_{\mathrm{CB}}$. This is one reason that Chao and
Bunge set $m=10$ (as noted above). In fact,~$\hat{N}_{\mathrm{CB}}$ fails
roughly as often as $\hat{N}_{\mathrm{ML}}$, although not necessarily in the
same situations.
\item\textit{Chao.} Chao (\citeyear{Chao1987,Chao1989}) proposed the nonparametric statistic
\[
\hat{N}_{\mathrm{Ch}}= n + \frac{f_1^2}{2f_2},
\]
which is valid as a (nonparametric) lower bound for $N$; we compute it
here as a benchmark. Note that $m\equiv2$.
\end{itemize}

We are currently investigating the asymptotic behavior of our
estimator~$\hat{N}$ in detail. Here we can make the following observations.
First, assume that the upper frequency cutoff~$m$ is selected as $m =
\max\{j\dvtx f_i>0,i=1,\ldots,j\}$, so that~$m$ is a random variable. For
the \textit{unweighted} case, that is, $\mathbf{W} = I_{m-1}$ in
Section \ref{ls}
above, it may be readily shown that $\hat{N}/N\rightarrow1$ in
probability as $N\rightarrow\infty$, when either $\mathbf{Y} =
[(i+1)f_{i+1}/f_i]$ and $(i+1)p_{i+1}/p_i = \gamma+ \delta i$
(the Katz condition), or $\mathbf{Y} = [\log((i+1)f_{i+1}/f_i)]$ and $\log((i+1)p_{i+1}/p_i) = \gamma+
\delta i$. If $\mathbf{W}= ({\operatorname{cov}}(\mathbf{Y}))^{-1}$ or a
diagonal matrix with positive variances as entries (similar to those
discussed in Section \ref{ls}), then we conjecture that analogous results can
be obtained (here $\mathbf{W}$ must be a function of $m$). The
convergence question is more complex for a weight matrix~$ \mathbf{\hat
W}$ that is estimated and perhaps approximated further (as in Section \ref{ls}),
although we believe that a~Slutsky-type argument will again yield
the desired consistency result. In any case, we note again that from
our practical experience a weighted estimator [even with estimated
weights using (\ref{dia})] increases the efficiency and reduces the
bias of the estimator considerably compared to the unweighted one (cf.
Table \ref{table7}).

\subsection{Data analyses}
We applied the proposed regression method and the alternative
procedures to the five data sets discussed above. The
results are shown in Table \ref{tab:resultstable}.
%
%
\begin{table}[b]
\vspace*{-6pt}
\caption{Data analyses. $\hat{N} = {}$weighted linear regression model;
$\hat{N}_{\mathrm{ML}} = {}$negative binomial maximum likelihood estimate; $\hat
{N}_{\mathrm{CB}} ={}$Chao--Bunge estimator; $\hat{N}_{\mathrm{Ch}} = {}$Chao lower bound;
SE${} = {}$standard error; $p = p$-value from $\chi^2$ goodness-of-fit
test; *${} = {}$estimation failed}\label{tab:resultstable}
\begin{tabular*}{\textwidth}{@{\extracolsep{\fill}}p{36pt}d{5.0}d{5.2}ccccccd{4.0}@{}}
\hline
\textbf{Study} & \multicolumn{1}{c}{$\bolds{\hat{N}}$} & \multicolumn{1}{c}{\textbf{SE}} & \multicolumn{1}{c}{$\bolds{p}$} &
\multicolumn{1}{c}{$\bolds{\hat{N}_{\mathrm{ML}}}$} & \multicolumn{1}{c}{\textbf{SE}} &
\multicolumn{1}{c}{$\bolds{p}$} & \multicolumn{1}{c}{$\bolds{\hat
{N}_{\mathrm{CB}}}$} & \multicolumn{1}{c}{\textbf{SE}} & \multicolumn{1}{c@{}}{$\bolds{\hat{N}_{\mathrm{Ch}}}$} \\
\hline
Meth. & 61\mbox{,}133 & 17\mbox{,}088.8 & 0.000 & * & * & * & * & * & 33\mbox{,}090\\
Polyps---low & 495 & 37.15 & 0.340 & 892 & 342.3 & 0.619 & 668 & 141.4
& 458\\
Polyps---high & 513 & 52.0 & 0.001 & 587 & \phantom{0}77.2 & 0.010 & 584 & \phantom{0}72.0 &
511 \\
Scrapie & 459 & 112.0 & 0.298 & * & * & * & * & * & 353 \\
Butterflies & 746 & 24.6 & 0.200 & 715 & \phantom{0}19.9 & 0.000 & 757 & \phantom{0}32.4 &
714\\
Microbial & 183 & 35.9 & 0.000 & * & * & * & * & * & 212\\
\hline
\end{tabular*}
\end{table}
Here the cutoff $m$ was selected for the weighted linear regression
model by taking the first $m$ at which $f_m>0$ and $f_{m+1}=0$; for the
ML procedure $m$ was selected by a goodness-of-fit criterion described
in Bunge and Barger (\citeyear{Bunge2008}), and $m\equiv10$ for $\hat{N}_{\mathrm{CB}}$ and
$\hat{N}_{\mathrm{Ch}}$.

\begin{figure}

\includegraphics{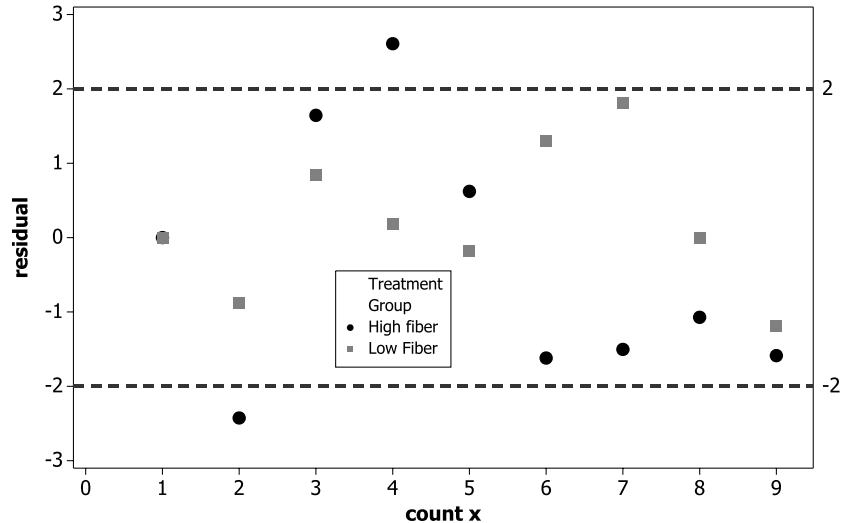}

\caption{Residual plot $(f_x-\hat f_x)/\sqrt{\hat f_x}$
versus $x$ for both treatment groups in the adenomatous polyps data set.}
\label{fig:residplot}
\vspace*{-6pt}
\end{figure}

We observe first that $\hat{N}$ gives an answer in every case, unlike
$\hat{N}_{\mathrm{ML}}$ and~$\hat{N}_{\mathrm{CB}}$. For the methamphetamine data,
although the $\chi^2$ $p$-value is low, the result appears reasonable,
especially with reference to the Chao lower bound. For the polyps---low
data, $\hat{N}$ gives the most precise result, with good fit; for
the polyps---high data, the same is true but with less good fit.
Despite the goodness-of-fit test in the latter case, though, residuals
plots for both polyps data sets indicate reasonable conformity with the
linear model, as shown in Figure \ref{fig:residplot}.
%
%
For the scrapie data it is interesting to note that $\hat{N}$ gives a
reasonable result with good fit while both $\hat{N}_{\mathrm{ML}}$\vspace*{1pt} and $\hat
{N}_{\mathrm{CB}}$ fail. For the butterfly data, $\hat{N}$ is comparable to $\hat
{N}_{\mathrm{CB}}$, with good fit of the linear model, while the ML result is
only slightly above the lower bound, with poor fit, indicating
difficulty with the ML\vspace*{1pt} numerical search. Finally, for the microbial
data, both $\hat{N}_{\mathrm{ML}}$ and $\hat{N}_{\mathrm{CB}}$ fail, while $\hat{N}<\hat
{N}_{\mathrm{Ch}}$ with poor fit, signaling that the data set is anomalous in
some way (in fact, it is highly skewed left). Overall, the weighted
linear regression approach shows up well in contrast to its competitors
for the negative binomial model.

\subsection{Discussion}
The main challenge in population-size estimation is arguably
heterogeneity, that is, the fact that in real applications the capture
probabilities or sampling intensities of the population units are not
all equal. The statistician must account for this in some way or risk
the severe downward bias of procedures based on the assumption of
homogeneity, that is, on ``pure'' binomial or Poisson models. Since the
time of Fisher, Corbet and Williams (\citeyear{Fisher1943}), considerable success has
been achieved using mixed-Poisson models with various mixture
distributions intended to model heterogeneity, including the gamma,
lognormal, inverse Gaussian, Pareto, generalized inverse Gaussian and,
more recently, finite mixtures of point masses or of exponentials
[Bunge and Barger (\citeyear{Bunge2008}), Quince, Curtis and Sloan (\citeyear{Quince2008}), B\"{o}hning
and Sch\"{o}n (\citeyear{Boehning20051})].\vadjust{\goodbreak} But the substantive applications, such as those
described in our examples here, typically do not offer a theoretical
basis for selection of a mixing distribution, so researchers have had
to search ever further afield for flexible and adaptable heterogeneity
models. This is partly due to a perception that the ``classical''
gamma-mixture or negative binomial model is too restrictive and
difficult to fit, both statistically and numerically.

However, existing mixed-Poisson-based procedures, whether frequentist
or Bayesian, are almost all based on the likelihood of the frequency
count data. Here we take a completely different perspective based on
the Katz relationship (\ref{katz}), finding that in many cases the
ratio of successive frequency counts $\hat{r}(x) = (x+1)f_{x+1}/f_x$
appears as an approximately linear function of $x$. This relationship
holds exactly for the gamma-mixture or negative binomial, and provides
an improved method both for fitting that model and for assessing its
fit. Furthermore, from the data-analysis perspective, the linear
relationship seems to hold across a wide variety of data sets; and from
the theoretical perspective, we know that every mixed-Poisson has (at
least) monotone increasing Katz ratios, and that the Katz distribution
family itself admits extensions in several directions. We therefore
believe that this perspective---looking at the data via $\hat{r}(x)$---opens
up a new method of applying the negative binomial model to
data, and that it gives us a view of a new and little-known territory
for exploring the robustness and extensions of that model.

\begin{appendix}\label{app}

\section*{Appendix: Simulation study, standard errors and dependence on the
truncation point}
\subsection{Comparative simulation study}
We begin with one further extension. The suggested weighted linear
regression estimator $\hat{N}$ depends on a~first-order Taylor
approximation which might not be good for larger values of $x$. One
might consider a second-order approximation, but this leads to an
estimator with large variance due to the functional relationship of~$x$
and $x^2$. An alternative linear approximation is possible by
developing $\log(k+x)=\log((k-1)+(x+1))$ linearly around $x+1$, leading
to the approximation
\[
\log(x+1) + (k-1)/(x+1)
\]
and the regression model
%
%
\begin{equation}
\log\biggl(\frac{(x+1)f_{x+1}}{f_x}\biggr)- \log(x+1) = \gamma^\prime
+ \delta^\prime/(x+1) + \varepsilon_x .
\label{model_2}
\end{equation}
We call this the \textit{hyperbolic model} (HM). The hyperbolic model is
also of very simple structure and prediction is possible since the
model is defined for $x=0$ leading to $\hat f_0 =f_1/\exp(\hat\gamma
^\prime+\hat\delta^\prime)$. We denote the estimator based on this
model by~$\hat{N}_{\mathrm{HM}}$.

\begin{table}
\caption{RMSE and Bias for estimators based upon the WLRM, the HM,
the Chao--Bunge estimator and the lower bound estimator of Chao,
$N=100$ and $N=1000$, $k=1, 2, 4, 6, 10$, where $k$ is the dispersion
parameter of the negative-binomial with mean $\mu=1$. Chao--Bunge
estimates have been computed only for positive values}\label{table12}\vspace*{4pt}
\vspace*{-4pt}
\begin{tabular*}{\textwidth}{@{\extracolsep{\fill}}ld{3.2}d{3.2}d{4.2}d{3.2}@{}}
\hline
\multicolumn{1}{@{}l}{$\bolds{k}$} & \multicolumn{1}{c}{\textbf{WLRM}} & \multicolumn{1}{c}{\textbf{HM}} & \multicolumn{1}{c}{\textbf{Chao--Bunge}} &
\multicolumn{1}{c@{}}{\textbf{Chao}}\\
\hline
\multicolumn{5}{@{}l}{RMSE $N=100$} \\
\phantom{0}1 & 25.36 & 366.89 & 1475.91 &27.60 \\
\phantom{0}2 & 31.93 & 816.54& 1145.43 &21.14 \\
\phantom{0}4 & 37.93 & 557.87& 585.20 &18.59\\
\phantom{0}6 & 43.56& 800.57 & 642.57& 18.21\\
10 &54.72 & 3453.55 &256.71 & 18.47\\
[3pt]
\multicolumn{5}{@{}l}{BIAS $N=100$} \\
\phantom{0}1 & -10.03& 115.98 &81.08 & -21.33\\
\phantom{0}2 & 4.39 & 124.90 & 52.11& -11.49\\
\phantom{0}4 & 12.22 &113.29 & 31.37 & -4.89\\
\phantom{0}6 & 15.23 & 116.89& 30.60& -2.07\\
10 & 16.93 & 162.21 & 17.01&-0.30\\[3pt]
\multicolumn{5}{@{}l}{RMSE $N=1000$} \\
\phantom{0}1 & 185.62 & 247.96& 191.25 &251.28 \\
\phantom{0}2 & 87.11& 206.02& 117.80& 152.88\\
\phantom{0}4 & 72.79 & 176.69& 96.55& 93.04\\
\phantom{0}6 & 75.81& 165.98 &86.61 &73.10\\
10 & 79.26 & 161.73& 81.08&59.70 \\
$\mu=0.5, k=0.5$&375.72&576.80&5247.90&471.19 \\[3pt]
\multicolumn{5}{@{}l}{BIAS $N=1000$} \\
\phantom{0}1 & -177.89 & 92.68& 23.70 & -247.25\\
\phantom{0}2 & -59.9 & 49.46 & 12.88&-145.51 \\
\phantom{0}4 & -1.88& -12.05& 9.96&-78.53 \\
\phantom{0}6 & 13.26 & -42.45&7.96 &-52.99 \\
10 & 21.88 &-72.31 &7.28 &-31.75\\
$\mu=0.5, k=0.5$&-368.16&
192.00&
-145.47&
-468.33\\
\hline
\end{tabular*}
\vspace*{-4pt}
\end{table}
%

%
%
\begin{table}[b]
\tablewidth=230pt
\caption{Estimated [using (\protect\ref{variance})] and true standard error
for WLRM estimator $\hat N$; $N=100$ and
$N=1000$, $k=1,2,4,6,10$, $\mu=1$; results are based on 10,000
replications}\label{table15}\vspace*{2pt}
\begin{tabular*}{230pt}{@{\extracolsep{\fill}}lcc@{}}
%
\hline
\multicolumn{1}{@{}l}{$\bolds{k}$} &$\bolds{\widehat{\mathit{S.E.}}(\hat N)}$& True $\bolds{\mathit{S.E.}(\hat N)}$ \\
\hline
\multicolumn{3}{@{}l}{$N=100$} \\
\phantom{0}1 & 26.94 & 23.06\\
\phantom{0}2 & 36.36 &30.00 \\
\phantom{0}4 & 44.23 & 38.02 \\
\phantom{0}6 & 44.13& 38.57 \\
10 & 41.88& 42.21 \\[3pt]
\multicolumn{3}{@{}l}{$N=1000$} \\
\phantom{0}1 & 52.31& 52.67 \\
\phantom{0}2 & 64.73&64.36 \\
\phantom{0}4 & 72.61& 71.64 \\
\phantom{0}6 & 75.68 & 73.51 \\
10 & 77.90&76.12 \\
\hline
\end{tabular*}
\end{table}

In the following simulation comparison, then, we compare $\hat{N}$,
$\hat{N}_{\mathrm{HM}}$, $\hat{N}_{\mathrm{CB}}$ and~$\hat{N}_{\mathrm{Ch}}$. We
generated counts from a negative binomial distribution\vadjust{\goodbreak} with
dispersion parameters equal to 1, 2, 4, 6 and 10 and event probability
parameter such that the associated mean matches 1. The population sizes
to be estimated were
$N=100$ and $N=1000$. For $N=1000$ a case with a
combination of $\mu=0.5, k=0.5$ was included which we have observed as
typical values in our data sets (Butterfly and Polyps data). A sample
$X_1,\ldots,X_N$ of size $N$ was generated from a negative binomial
distribution with parameters as described above and the associated
frequency distribution $f_0, f_1,\ldots, f_m$ was determined; then
$f_0$ was ignored and $f_1, \ldots, f_m$ were used to compute the
various estimators.
This process was repeated
1000 times and bias, variance and MSE were calculated from the
resulting values.
The results are shown in Table \ref{table12}. Clearly, $\hat{N}$
performs better than $\hat{N}_{\mathrm{HM}}$ since the former always has smaller
MSE than the latter. In fact, there are only three cases in which $\hat
{N}_{\mathrm{HM}}$ had smaller bias than $\hat{N}$, namely, $N=1000$ and $k=1,
2$ as well as the combination $\mu=0.5, k=0.5$, and the smaller bias
here was balanced by the smaller variance of $\hat{N}$. Hence, we do
not consider $\hat{N}_{\mathrm{HM}}$ any further. We see in addition that $\hat
{N}$ and $\hat{N}_{\mathrm{CB}}$ overestimate the true size $N=100$, whereas
$\hat{N}_{\mathrm{Ch}}$ tends to
underestimate. We need to point out that $\hat{N}_{\mathrm{CB}}$ produced many
negative values, so its bias and RMSE were evaluated on the basis of
the positive values. The bias of $\hat{N}$ is smaller than that of $\hat
{N}_{\mathrm{CB}}$ for $N=100$, although this reverses for $N=1000$, and the
bias is of the same size as that of $\hat{N}_{\mathrm{Ch}}$ for $N=100$ and
becoming smaller for $N=1000$. Also, the RMSE of $\hat{N}_{\mathrm{CB}}$ is a
lot larger than that of $\hat{N}$.
The situation changes for $N=1000$. In this case both the bias and MSE
for $\hat{N}$ are lower than those from
$\hat{N}_{\mathrm{Ch}}$ for every value $k$ of the
dispersion parameter. We notice, however, that $\hat{N}_{\mathrm{CB}}$ shows a
reduced bias, but the RMSE of $\hat{N}$ is still smaller.
Overall, we find that $\hat{N}$ and $\hat{N}_{\mathrm{CB}}$ are behaving
somewhat similarly for larger population sizes; however, a major
benefit of $\hat{N}$ is that it is well defined in the many situations
where $\hat{N}_{\mathrm{CB}}$ fails.

\subsection{Standard errors}

In Table \ref{table15} we compare the standard error calculated from
(\ref{variance}) with the true standard error. This was done by taking
$10\mbox{,}000$ replications of $\hat N$, say, $\hat N_i, i=1,\ldots, 10\mbox{,}000$.
Then the mean of $(1/10\mbox{,}000)\sum_i \widehat{\operatorname{Var}}(\hat N_i)$ was
computed and the root of it forms column 2 in Table \ref{table15}. The
third column was constructed by simply computing the empirical variance
of $\hat N_i, i=1,\ldots, 10\mbox{,}000$. We see that the approximation is good
(and always conservative)\vadjust{\goodbreak} for larger values of $N$ and reasonable for
smaller values of $N$.
Finally, we would like to mention the bootstrap as an alternative to
the approximate standard errors given above. The bootstrap is
straightforward to implement here: first obtain $\hat{N}$ from the
original data; then resample (simulate) $f_0^*, f_1^*, \ldots$ based on
the fitted $\hat{p_0}, \hat{p_1}, \ldots$; then delete~$f_0^*$ and
calculate a new $\hat{N}^*$ from the new sample. Replicate this
procedure $B$ times (say) and from the resulting $\hat{N}^*$'s
calculate a standard error for $\hat{N}$, percentile-based confidence
intervals, and so forth.

%
\begin{table}
\caption{Dependence of the weighted least-squares $\hat{N}$ and the
Chao--Bunge estimator on the truncation point, compared for all data sets}
\label{table11a}
\vspace*{-6pt}
\begin{tabular*}{\textwidth}{@{\extracolsep{\fill}}lcd{5.0}cd{4.0}cccd{4.0}@{}}
\hline
& \multicolumn{2}{c}{\textbf{Polyps--low}} & \multicolumn{2}{c}{\textbf{Polyps--hi}} & \multicolumn{2}{c}{\textbf{Butterflies}} & \multicolumn{2}{c@{}}{\textbf{Microbial}}
\\[-6pt]
& \multicolumn{2}{c}{\hrulefill} & \multicolumn{2}{c}{\hrulefill} & \multicolumn{2}{c}{\hrulefill} & \multicolumn{2}{c@{}}{\hrulefill}
\\
\multicolumn{1}{@{}l}{$\bolds{m}$} & \textbf{WLRM} & \textbf{C--B} & \textbf{WLRM} & \textbf{C--B} & \textbf{WLRM} & \textbf{C--B} & \textbf{WLRM} & \textbf{C--B} \\
\hline
\phantom{0}3 & 609 & 411 & 881 & 446 & 754 & 682 & 767 & 266 \\
\phantom{0}4 & 525 & 440 & 620 & 459 & 744 & 696 & 364 & 492 \\
\phantom{0}5 & 509 & 471 & 542 & 472 & 776 & 715 & 364 & 492 \\
\phantom{0}6 & 523 & 524 & 513 & 482 & 759 & 727 & 364 & -240 \\
\phantom{0}7 & 519 & 596 & 512 & 497 & 752 & 737 & 364 & -240 \\
\phantom{0}8 & 503 & 643 & 519 & 532 & 746 & 746 & 364 & -75 \\
\phantom{0}9 & 495 & 668 & 510 & 570 & 741 & 752 & 216 & -59 \\
10 & 495 & 668 & 510 & 570 & 732 & 757 & 212 & -49 \\
11 & 495 & 844 & 510 & 586 & 726 & 761 & 214 & -42 \\
12 & 495 & 844 & 506 & 607 & 724 & 765 & 205 & -43 \\
13 & 495 & 844 & 506 & 607 & 717 & 768 & 197 & -45 \\
14 & 495 & 844 & 506 & 607 & 718 & 774 & 195 & -46 \\
15 & 495 & 844 & 506 & 607 & 712 & 777 & 195 & -46 \\
16 & 495 & 844 & 506 & 607 & 711 & 783 & 195 & -46 \\
17 & 495 & 844 & 506 & 607 & 708 & 788 & 195 & -46 \\
18 & 495 & 844 & 506 & 607 & 704 & 792 & 182 & -48 \\
19 & 495 & 844 & 506 & 607 & 704 & 797 & 182 & -48 \\
20 & 495 & 844 & 506 & 607 & 701 & 802 & 182 & -48 \\
21 & 495 & 844 & 506 & 607 & 698 & 805 & 182 & -48 \\
22 & 495 & 1821 & 506 & 607 & 695 & 807 & 182 & -48 \\
23 & 495 & 1821 & 506 & 607 & 693 & 808 & 182 & -48 \\
24 & 495 & 1821 & 506 & 607 & 692 & 810 & 182 & -48 \\
28 & 495 & -2250 & 506 & 607 & & & 182 & -48 \\
29 & & & 506 & 607 & & & 182 & -43 \\
31 & & & 506 & 1063 & & & 182 & -43 \\
42 & & & 506 & 1063 & & & 182 & -33 \\
53 & & & 506 & 1063 & & & 182 & -27 \\
77 & & & 506 & -301 & & & & \\
\hline
\end{tabular*}
\end{table}

\subsection{Dependence of estimators on the truncation point}
Table \ref{table11a} shows the dependence of $\hat{N}$ vs. that of
$\hat{N}_{\mathrm{CB}}$ on the truncation point for the first four data sets
considered here. The behavior of $\hat{N}$ is notably more stable than
$\hat{N}_{\mathrm{CB}}$ in this regard, except perhaps for the butterfly data.
The negative binomial MLE and the coverage-based nonparametric
estimators also display considerable instability with respect to $m$,
except in the case of the butterfly data (results not shown). The only
other procedure we know of that is relatively robust with respect to
$m$ is the parametric estimator based on finite mixtures of geometrics
(i.e., Poisson where the Poisson mean is distributed as a finite
mixture of exponentials); for details on this model see Bunge and
Barger (\citeyear{Bunge2008}).
\end{appendix}

\section*{Acknowledgments} We are grateful to the Editor and to a
referee for many valuable comments that helped to improve the
manuscript, touching on too many points to list here.
Some of this research was completed while D. B\"{o}hning was visiting the
Department of Statistical Sciences of Cornell University. D. B\"{o}hning would like
to thank Cornell University as well as the University of Reading for
supporting this visit. I. Rocchetti would like to thank Marco Alf\'{o} for
giving suggestions on a previous version of this manuscript.
%

%


\printaddresses

\end{document}